\documentclass[12pt,preprint]{aastex}
\usepackage{graphicx}

\begin{document}

\title{\bf The Discovery of a High Redshift X-Ray Emitting QSO
Very Close to the Nucleus of NGC 7319}

\author{Pasquale Galianni\altaffilmark{1}, E. M.
Burbidge\altaffilmark{2}, H. Arp\altaffilmark{3},\\ V. Junkkarinen\altaffilmark{2}, G.
Burbidge\altaffilmark{2}, Stefano Zibetti\altaffilmark{3}}

\altaffiltext{1}{Dipartimento di Fisica' Universit\`a di Lecce, Via Arnesano I - 73100
Lecce, Italy} \altaffiltext{2}{Center for Astrophysics and Space Sciences 0424,
University of California, San Diego, CA 92093, USA} \altaffiltext{3}{Max-Planck-Institut
f\"ur Astrophysik, 85741 Garching, Germany}


\begin{abstract}
A strong X-ray source only 8\arcsec~ from the nucleus of the Sy2 galaxy NGC 7319 in
Stephan's Quintet has been discovered by Chandra.  We have identified the optical
counterpart and show that it is a QSO with z$_e$ = 2.11{\bf 4}. It is also a ULX with
L$_x$ = 1.5 x 10$^{40}$ erg sec$^{-1}$. From the optical spectra of the QSO and
interstellar gas of NGC 7319 at z = .022 we show that it is very likely that the QSO is
interacting with the interstellar gas.
\end{abstract}

\keywords{galaxies: active -­ galaxies:individual (NGC 7319) -­ quasars: general ­- X-ray
sources}

\section{INTRODUCTION}
In the last few years observations from Chandra and XMM-Newton have shown that there are
many discrete, powerful X-ray-emitting sources which lie close to the nuclei of spiral
galaxies, often, apparently inside the main body of the galaxy (Foschini et al. 2002a,
2002b; Pakull and Mirioni 2002; Roberts et al. 2001; Goad et al. 2002).  Typical
separations are from $\sim$ 1\arcmin~ to 5\arcmin.

Since they are emitting at power levels above about 10$^{38.5}$ erg sec$^{-1}$, they
cannot be normal X-ray binaries.  They have been called ULX (ultraluminous X-ray sources)
or IXO, intermediate luminosity X-ray sources (Colbert and Ptak 2002).  It has been
concluded that they are either binary systems with black hole masses in the range 10$^2 -
10^4$ M$_\odot$, or they are X-ray emitting QSOs.  Last year Burbidge et al. (2003)
suggested that they were likely to be QSOs with a wide range of redshifts. If this is the
case, the fact that they are all very close to the centers of the galaxies suggests
strongly that they are physically associated with these galaxies and are in the process
of being ejected from them.  This is a natural conclusion following from the earlier
studies by Radecke (1997) and Arp (1997), who showed that there is a strong tendency for
QSOs to cluster about active spiral galaxies.  Many cases of this kind have been found
(eg. near the AGN galaxies NGC 1068, 2639, 3516, 3628).  The typical separations $\Delta
\theta$ between these QSOs and the galaxies in these cases are $\sim$ 15\arcmin~ -
20\arcmin.  It is clear that if the separations are smaller than this, as is the case in
general for the ULXs, there will be an even greater likelihood that the QSOs and galaxies
are physically associated.

Colbert and Ptak (2002) gave a list of 87 IXOs which lie within 5\arcmin~ of the centers
of galaxies.  Recent spectroscopic studies (Arp et al. 2004) have shown that three in
this list plus 21 other cases which fit these IXO criteria are already known to be normal
QSO's.

In this paper we report on the investigation of a new Chandra ULX optically identified by
one of us (Galianni 2003). It lies only 8\arcsec~ from the nucleus of the Sy2 galaxy NGC
7319, a member of Stephan's Quintet.

In the following sections we give details of the identification of
the optical object and analysis of its spectrum.

We then discuss the spectroscopic and direct optical evidence
relating to its connections with the galaxy.  Finally we summarize
our result and its bearing on the general problem of ULXs.

\section{IDENTIFICATION OF THE X-RAY SOURCE}

In Fig. 1 we show that there are only two optical objects in the field around the nucleus
of NGC 7319. Both are to the south and one resembles a luminous gas cloud and the other
appears stellar. The X-ray position given in Trinchieri et al. (2003) for source No. 9 in
their Table 1 lies very close to the stellar object. Chandra X-ray positions generally
have absolute accuracies of around 1 arcsec. But since there are no X-ray point source
calibrators in the field it was necessary to check the zero point of the X-ray
coordinates by another method. This was achieved by one of us (Galianni) by superposing
the Chandra X-ray contours on that part of the field containing both the stellar object
and the Seyfert nucleus. Fig. 2 demonstrates that by moving the X-ray position about one
arcsec east, the X-ray contours of the nucleus are superposed on the central part of the
nucleus, and at the same time the X-rays from the ULX coincide exactly with the HST image
of the stellar object. Thus we have identified the ULX with the marked stellar object
which has the following properties:

R.A.(2000) = 22$^h:36^m:03.71^s$  Dec.(2000) = 33$^0:58\arcmin:24\arcsec.06$

 V = 21.79 ~~  B-V = .43 ~~ V-I = .98.

\section{THE SPECTRUM OF THE ULX}

The optical spectrum of the ULX was obtained on the night of 2 October 2003, with the Low
Resolution Imaging Spectrograph (LRIS) (Oke, Cohen et al. 1995) attached to the Keck I
10m telescope on Mauna Kea.  We show the spectrum in Fig. 3, and in Table 1 we show the
emission lines and their identifications.  The ULX is clearly a QSO with an emission
redshift z = 2.11.

The emission lines in this QSO are broad, and are chopped up by absorption so their
wavelengths have been estimated only to the nearest angstrom. However, if we average the
redshifts of the five strongest lines in Table 1 we obtain z = 2.114. This is an
interesting result because there are sharp absorption lines near the centers of the
Ly$\alpha$, Si IV, and C IV emission lines with redshifts of 2.117, 2.110 and 2.121
respectively. This suggests that these major emission lines are self absorbed - a
possibility which will be commented on further in the discussion.

\begin{table}[h]
\caption{Emission lines measured} \label{Table1} \vspace{0.3cm}
\begin{tabular}{lccc}
Line & $\lambda_0$ & obs  & z\\
& & \\
Ly$\alpha$ & 1215.6 & 3791 & 2.12\\
Si IV - O IV] & 1398.5 & 4335 & (2.10)\\
C IV & 1549.1 & 4840 & 2.12\\
C III] & 1908.7 & 5942 & 2.11\\
Mg II & 2798.8 & 8733 & 2.12 \\
\end{tabular}
\end{table}

\section{EMISSION LINES FROM GAS IN NGC 7319 NEAR THE QSO}

In obtaining the spectrum the candidate QSO was placed at the center of a 0.7\arcsec~
slit which was oriented at a position angle of p.a. = 205 deg. on the sky. (See Fig.4).
The two-dimensional spectra in Fig 5 show that the slit passes over the HII region about
2.4\arcsec~ to the SSW of the main target. On the other side about 4.6\arcsec~ to the NNE
the slit passes over a bright part of the disk of NGC 7319. For an extent of about
8\arcsec~ along the slit we see the strongest emission lines from the galaxy (see color
picture Fig. 4 for the appearance of these regions.) Further out (shown only on the red
spectrum) the slit appears to pass over another HII region. But the major emission lines
from the galaxy are found to be coming from the bright disk-like region of the galaxy
just south of the nucleus as shown in Figs. 1 and 4.

In Table 2 we show a list of the emission lines from gas in the galaxy measured in the
red spectral region, plus the lines of $[OII]$ and $[OIII]$, also from the galaxy. The
redshift (averaged) is z = 0.0224, and this can be compared with the redshift of the
centroid of this galaxy, z = 0.0225 (NED). Thus the gas in this region has a redshift in
agreement with, and within 30 km/sec, of the centroid of NGC 7319.

\begin{table}[h]
\caption{Emission lines due to galactic gas of NGC 7319 4.6\arcsec~ to NNE of the QSO}
\label{Table2} \vspace{0.3cm}
\begin{tabular}{lccc}
Line & $\lambda_0(\AA)$ & obs$(\AA)$ & z\\
& & \\
$[OII]$ & 3727.3 & 3808.2 & 0.0217\\
$[OIII]$ & 4958.91 & 5063.4 & (0.0211)\\
$[OIII]$ & 5006.84 & 5114.1 & 0.0214\\
$[OI]$  &  6300.23 & 6437.9 & 0.0219\\
$[NII]$ & 6549.85 & 6696.59 & 0.0224\\
$H\alpha$  & 6564.61 & 6711.47 & 0.0224\\
$[NII]$ & 6585.28 & 6732.63 & 0.0224\\
$[SII]$ & 6718.29 & 6868.17 & 0.0223\\
$[SII]$ & 6732.67 & 6883.17 & 0.0224 \\
\end{tabular}
\end{table}

The emission lines in the galaxy due to $[OII]$ and $[OIII]$ in the blue, however, have a
different shape and extent from those in the red and are blue shifted, on average, by
about 200 - 300 km/sec.

\subsection{Forbidden $[OII]$ Emission From The Gas In NGC 7319}

In the 2d spectra shown in Fig. 5 it is clear that the emission lines in that region of
the galaxy projected NNE of the QSO show the forbidden lines to be stronger than the
permitted lines. In the bottom spectrum of Fig. 5, which covers the spectrum from
6400$\AA$ to 7000$\AA$, one can see the expected spectrum of an HII region with prominent
H$\alpha$. This is also true of the spectrum of the HII region about 3.2\arcsec~ to the
SSW. In contrast, the lines in the galaxy are very strong in the forbidden transitions of
nitrogen and sulfur and weak in H$\alpha$. This is typical of the strong star burst
regions in a Seyfert disk as compared to isolated HII regions (Bransford et al. 1998). In
Fig. 5 center, $[OIII] \lambda$ 5007 is shown to be very strong and misshapen but H$
\beta$ is not visible at all.

The enormous strength and extent of the $[OII] \lambda$ 3727 emission near the projected
position of the QSO is particularly interesting. (The equivalent width (eqw) = 26$\AA$.)
In the 2-dimensional spectrum shown in Fig. 5 the $[OII]$ emission is about 120 pixels
(16 arc sec) long. This is about twice the extent of the galaxy emission lines in the
red, and it indicates that the line comes from a different, larger region. In fact it
essentially samples the full extent of the inner disk as pictured in Fig. 4. The QSO is
also better centered on this line, which then extends also SSW of the QSO.

Empirically we know that the $[OII]\lambda$ 3727 doublet is characteristic of the outer
low density regions of galaxies where the metastable state of $[OII]$ is not
collisionally de-excited. But to excite this large population of excited $[OII]$ atoms in
a low density environment requires a powerful source of ionizing radiation.

In a detailed study of the extended emission line region(EELR) in NGC 7319 Aoki et al.
(1996) showed that the ionizing photon flux from the line-of-sight UV/X-ray flux is
insufficient to explain the $H\alpha$ emission line luminosity.  They showed that the
photon deficit is model dependent but the possible range of values is from $>$ 5 to $>$
460. They therefore concluded that the radiation field is strongly anistropic.

However, they did not consider the possibility that a secondary source could be
responsible for the anomaly.  Despite the fact that the ULX-QSO has a much smaller X-ray
count rate than the central nucleus it seems very likely that this enormously strong
$[OII]$ emission near the projected position of the QSO is due to the close proximity of
this secondary source of UV/X-ray flux, i.e. the QSO is interacting strongly with the
interstellar gas in the disk of the galaxy.

One possibility is that the QSO emerges from the nucleus of NGC
7319 with a component of velocity toward the observer perhaps
coming slightly out of the plane on the observer's side. The QSO
then excites the lower density gas in this region (or has
entrained some gas from the regions it has recently passed
through).

\subsection{$[OI]$ Emission}

It is interesting in this connection to note the presence of the $[OI]$ 6300${\AA}$ line
which we measure in the galaxy disk adjoining the QSO. This line is typical of inner
Seyfert regions as shown by Bransford et al. (1998). They remark `` . . . we present
evidence for vigorous, compact star formation enclosed by very extended $[OI]$ lambda
6300 emission, suggestive of the boundary between a diffuse outflow and the surrounding
interstellar medium".  They then suggest that $[OI]$ emission is caused by ``weak (150
km/sec) shocks in the circumnuclear environment (Dopita and Sutherland 1995)".
Intriguingly we derive for our Seyfert disk between the galaxy nucleus and the QSO a
redshift of z = 0.0224 - and a redshift of z = 0.0219 for our $[OI]$ emission. The
difference is just 150 km/sec (estimated uncertainty ${\pm}$ 30 km/sec).

\subsection{Evidence For Outflow In The Direction Of The QSO}

Our one long slit placement across the ULX/QSO with the Keck LRIS instrument confirms
several of the main results of Aoki et al. (1996) as far as outflow is concerned.  For
example they concluded that there is a gaseous ``high velocity and large scale outflow
into the extended emission line region (EELR) toward the south-west direction"
``coincident with the direction of the radio emission" ``The velocity of the outflow
comes up to 500km/sec."

In our spectrum the slit passed over the EELR to the south of the Seyfert nucleus.
Assuming that the systemic redshift of the galaxy is z = 0.0225, Table 1 shows $[OII]$ at
-210 km/sec, $[OIII]$ at -300 km/sec and the weaker $[OIII]$ line at -390 km/sec. Thus
the outflow of the gas is very high.

\subsection{The Wavelength Coincidence of the $[OII]$ emission line in NGC 7319\\
with Ly$\alpha$ in the QSO}

It is remarkable that the $[OII]$ line in the 2d spectrum of Fig. 5 closely intersects
the Ly$\alpha$ line in the QSO. Figure 6 is introduced here in order to show the
exactness of this relation. The $[OII]$ line is about $7 \AA$ (FWHM) and reaches a peak
of about 3 x 10$^{-16}$ erg cm$^{-1}$ sec$^{-2} \AA^{-1}$. The Ly$\alpha$ line width is
about $150 \AA$ (FWHM) and peaks at about 3 x 10$^{-17}$ in the same flux units. It is
clear that the major QSO line as well as its continuum is well placed to pump the excited
state of $[OII]$ emission.

It is also true that the $7 \AA$ (FWHM) of the $[OII]$ line gives a velocity width of 550
km/sec. (The $[OIII]$ 5007 line width gives 540 km/sec). This would imply that the
passage of the QSO through the material of the galaxy has produced a dynamical
disturbance as well as a radiative excitation.

The color picture Fig. 4 gives pictorial evidence in support of a model where the QSO has
been ejected from the nucleus of the Seyfert NGC 7319. It is seen that there is a
luminous connection reaching from the nucleus (just at the top of the picture frame) down
in the direction of the ULX/quasar, stopping about 3\arcsec~ from it. It is also apparent
that this connection or wake is bluer than the body of the galaxy.

\section{ABSORPTION IN THE QSO SPECTRUM FROM THE
DISK OF NGC 7319?}

There are no signs of background objects showing through the disk in this HST picture of
the inner regions of NGC 7319 (Fig.1). This accords with our expectation that the
absorption in the disk near the center of this Seyfert galaxy would block out any objects
behind it. Nevertheless, if the QSO were in the disk, or on the near side of the disk as
we observe it, it could still show some signs of absorption from interstellar gas in the
disk.

Taking the apparent redshift of NGC 7319 as z = 0.0225 gives expected wavelengths of
$\lambda\lambda$ = 4056.0${\AA}$ and 4020.4${\AA}$ respectively for interstellar H and K
in the disk of the Seyfert galaxy. In this wavelength interval of our Keck QSO spectrum
there are two lines, possibly significantly above the noise, at 4055.07 and 4019.31,
giving z = 0.0220 and z = 0.0218. If these are real they have eqw = 1.1 and 0.8$\AA$.

However, two deep, narrow absorption lines (eqw $\sim$ 1.9 and 2.3$\AA$) are visible at
$\lambda_{obs} = 6017.3$ and 6023.1. They can be identified with Na I 5890.0 and 5895.9
at z = 0.02157 and z = 0.02161. They are separated by $5.8{\AA}$ and clearly represent
absorption by the sodium doublet in the gas of the interstellar medium. Their mean
redshift of z = 0.0216 is -240 km/sec less than the emission line redshifts in the red, z
= 0.0224, from Table 2. Perhaps this gas is associated with the $[OII]$ and $[OIII]$ gas
discussed in section 4 as having redshifts of the order of -200 to -300 km/sec relative
to the center of the galaxy.

In table 3 we give the redshifts of major stellar absorption features in this region of
NGC 7319.

\begin{table}[h]
\caption{Absorption lines in the galaxy 4.6\arcsec~ to NNE of QSO} \label{Table3}
\vspace{0.3cm}
\begin{tabular}{lccc}
Line & $\lambda_0$ & obs & z\\
& & \\
K       & 3933.7 & 4018.26 & 0.02158\\
H       & 3968.5 & 4054.8  &  0.02175\\
Gband & 4304.4 & 4400.45 & 0.0223\\
Na D  &  5892.5 & 6021.8 & 0.0219\\
\end{tabular}
\end{table}

\subsection{Is The QSO Behind The Galaxy?}

It is not surprising that interstellar sodium $D_1$ and $D_2$ absorption are seen in the
spectrum of the QSO. If the ejected gas and the QSO lie near the plane of the disk,
however disrupted that may be, we would still expect about half the possible optical
depth of gas between the QSO and the observer. But we have no way of knowing whether the
amount of gas observed here represents the total column of gas through the system, half,
or even less.

One obvious question suggests itself, namely: Does the color of the QSO indicate that it
is inordinately reddened and therefore obscured as if it were a background object?  Of
course that would require smooth conditions in the galaxy and a precise value for the
unreddened color. But we can make an empirical test by selecting 32 QSOs in a large
sample region of the Hewitt-Burbidge Catalog (Hewitt \& Burbidge 1993) which have
redshifts $2.0 \leq z \leq 2.2$. The average redshift is z = 2.09 and the average
(B-V)$_{ave} = 0.26 \pm .18$ (mean deviation). So we see the measured B-V = 0.43 for the
QSO is somewhat reddened but within the average deviation.

But if we compare the B-V of this ULX with fainter apparent magnitude QSOs from the
Hewitt-Burbidge Catalogue we find that it is about 0.1 to 0.2 mag. bluer than average. It
is also true that the continuum of the QSO is approximately flat at a flux level about
10$^{-17}$ erg cm$^{-1}$ sec$^{-2} \AA^{-1}$ all the way into the far red.

\section{AN ACCIDENTAL SUPER POSITION?}

If QSOs which are point X-ray sources are randomly distributed the probability that a QSO
will lie $\theta$ arc minutes from any point is given by

p = 8.64 x 10$^{-4}~\theta^2~\Gamma$

\noindent
where $\Gamma$ is the surface density of the QSO/X-ray sources.

If $\theta$ = 10\arcsec,    p  = 2 x 10$^{-5}~\Gamma$.

The surface density of QSOs is well established to be about 5 - 20 per square degree down
to 20$^m$ and it begins to level off to about 50 per square degree at $21^m$.  As was
pointed out earlier many QSOs which are ULXs have already been identified (cf. Arp et al.
2004) and they lie at an average distance from the center of the galaxy at between 4 and
5 arc minutes.  The two closest, before this QSO was discovered, were the QSO near NGC
4319 ($\theta$ = 43\arcsec) and the QSO near NGC 4168 ($\theta$ = 45\arcsec).  But all of
these are much brighter optically than this QSO, the average apparent magnitude being
$19^m$.1.

Since this QSO is so much closer to the nucleus of NGC 7319 than any of the others the
fact that it is about $2^m$.5 fainter than the others would strongly suggest that, if it
is a background QSO, it appears to be much fainter because of absorption in the
foreground galaxy.  This means that in calculating the probability we should use the
surface density of the bright QSOs.  Thus if we put $\Gamma$ = 5,

p = 10$^{-4}$.

Even if we put $\Gamma$ = 50,   p = 10$^{-3}$.

There may very well be a higher surface density of point X-ray sources, which might mean
that they will turn out to be much fainter optical QSOs.  But at present there is no
general evidence for this.  Moreover, the fact that we have detected the QSO in NGC 7319
at m = 21$^m$.8 sets a limit as far as probability arguments are concerned.

The very low probability of a chance superposition is yet more evidence in favor of the
view that this QSO is at the distance of NGC 7319.

\section{CONCLUSIONS AND DISCUSSION}

We have clearly demonstrated that the ULX lying 8 arc sec from the nucleus of NGC 7319 is
a high redshift QSO.  This is to be added to a list of more than 20 ULX candidates which
have all turned out to be genuine QSOs (cf. Burbidge, Burbidge \& Arp 2003; Arp,
L\'{o}pez-Corredoira and Guti\'{e}rrez 2004).  Since all of these objects lie within a
few arc minutes or less of the centers of these galaxies, the probability that any of
them are QSOs at cosmological distance, observed through the disk of the galaxy, is
negligibly small.  Thus this is further direct evidence that high redshift QSOs are
generated and ejected in low redshift active galaxies. The case described here is
particularly interesting since there is a considerable amount of evidence that the QSO is
interacting with the gas in the main body of the galaxy.  Most of the evidence has been
discussed in the previous sections.

One of the most pertinent observations comes from the radio studies of this galaxy at
20cm by van der Hulst and Rots (1981). Radio emission from the Seyfert nucleus extends to
the SW, bends over to the south and is traceable to within 5\arcsec  of the QSO.  Aoki et
al. (1999) confirmed the existence of radio ejection from the nucleus and concluded that
emission features to the SW are ``from gas compressed by the bow shock driven by the
outward moving radio plasmoid".  Perhaps the QSO is that plasmoid.

In Figs. 2 and 4 we had independently seen their ``V-shaped
feature" and concluded that it represented a luminous extension or
wake trailing the QSO.  Their final 1999 conclusion was: ``We
interpret the V-shaped feature as emission \textit{from gas
compressed by the bow shock driven by the outwardly moving
plasma." We have already noted that our spectrum in this region
quantitatively agreed with their 1996 velocity of outflow measure
of up to -500 km/sec.}

As was discussed earlier Aoki et al. (1996) were concerned with the source of ionizing
photons in the gas. They found a large photon deficit.  However, with the QSO in the
center of the emission region we have a secondary source with M$_V \simeq$ -13.8. Thus we
have a source of Ly$\alpha$ pumping directly into a huge bubble of $[OII]$ emitting gas,
together with synchrotron emission.

Further X-ray radio and optical observations following this evidence for a powerful
discrete source -- the QSO, exciting the interstellar gas -- are needed. For example high
resolution direct imaging with narrow band filters could tell us a great deal.  Data with
higher S/N shortward of Ly$\alpha$ emission would be useful, since this is where
Ly$\alpha$ forest absorption is seen in many QSOs.  Our spectrum, very noisy here, does
not show such features.

This is the only system found so far in which there is the possibility of demonstrating
even more clearly that the QSO and galaxy are interacting.

\section{ACKNOWLEDGEMENT}

We would like to thank Jeremy Walsh ST-ECF for providing the magnitude and colors of the
candidate object from the ST exposures.

The authors wish to recognize and acknowledge the very significant
cultural role and reverence that the summit of Mauna Kea has
always had within the indigenous Hawaiian community.  We thank the
staff of the W. M. Keck Observatory for their expert assistance
during the night of 2 October, 2003.

\clearpage

\section{REFERENCES}

Aoki, K., Ohtani, H., Yoshida, M. \& Kosugi, G. 1996, AJ, 111, 140

Aoki, K., Kosugi, G., Wilson, A. \& Yoshida, M. 1999, ApJ, 521, 565

Arp, H. 1997, A\&A, 319, 33

Arp, H., L\'opez-Corredoira, M. \& Guti\'errez, C. M. 2004, A\&A, 418, 877

Bransford, M., Appelton, P., Heisler, C., Norris, R. \& Marston, A. 1998, ApJ, 497, 133

Burbidge, G., Burbidge, M. \& Arp, H. 2003, A\&A, 400, L17

Colbert, E. \& Ptak A. 2002, ApJS, 143, 25

Dopita, M. \& Sutherland, R. 1995, ApJ, 455, 468

Foschini, L. et al. 2002a, arXiv: astro-ph/0206418

Foschini, L. et al. 2002b, arXiv: astro-ph/0209298

Galianni, P. 2003, Coelum Feb 2004, 54

Goad, M. R., Roberts, T. P., Knigge, C. \& Lira, P. 2002, MNRAS, 335, L67

Hewitt, A. \& Burbidge, G. 1993, ApJS, 87, 451

Oke, J., Cohen, J. Carr, M. et al. 1995, PASP, 107, 375

Pakull, M. \& Mirioni, L. 2002, astro-ph/0202488

Radecke, H.-D. 1997, A\&A, 319, 18

Roberts, T. P. et al. 2001, MNRAS, 325, L7

Trinchieri, G., Sulentic, J., Breitschwerdt, D. \& Pietsch, W. 2003, A\&A, 401,173

van der Hulst, J. \& Rots, A. 1981, AJ, 86,1775

\clearpage

Fig. 1  ---  Multi color HST image of the central region of NGC
7319 showing the optical ULX candidate 8 arcsec south of the
nucleus.

Fig. 2  ---  Chandra X-ray isophotes with zero point correction which brings X-ray
sources onto center of NGC 7319 nucleus and onto optical candidate.

Fig. 3  --  Keck spectrum of NGC 7319 ULX/QSO. Ordinate gives intensity in erg cm$^{-1}$
sec$^{-2} \AA^{-1}$, the horizontal scale gives measured wavelength.  QSO lines at z =
2.11 from left are $Ly\alpha, SiIV-OIV], CIV$ and $CIII]$.

Fig. 4  --  Enhanced HST image showing filament or ``wake" moving down from the center of
NGC 7319 to within about 3.4\arcsec of the QSO. Processed photograph reproduced courtesy
``Coelum" Astronomia (Italia).

Fig. 5  --  2D spectra showing $Ly\alpha$ in the QSO and $[OII]\lambda$3727 in galaxy
(upper); the region around $H\beta$ and $[OIII]$ in galaxy (middle); and bottom,
$H\alpha$ and forbidden emission lines in the red (with smaller scale) for the galaxy.
The ordinate scale is in pixels.

Fig. 6  --  The heavily smoothed profile of the $Ly\alpha$ line in the QSO, showing where
the $[OII]\lambda$ 3727 doublet in the surrounding region of the galaxy appears at
$\lambda 3808.2\AA$.  The ordinate is flux in erg cm$^{-1}$ sec$^{-2}\AA^{-1}$, the
abscissa is measured wavelength.  The $[OII]$ FWHM = 7.2${\AA}$ close to the $Ly\alpha$
peak, but reaches a factor 10 times higher in intensity. The base of the \textit{[OII]}
line in the galaxy spectrum is about 5 x 10$^{-17}$ erg cm$^{-1}$ sec$^{-2}\AA^{-1}$.

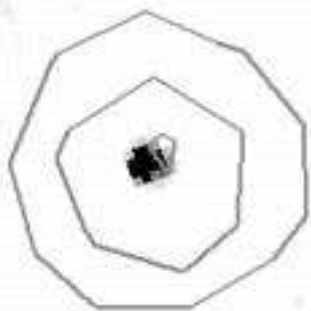
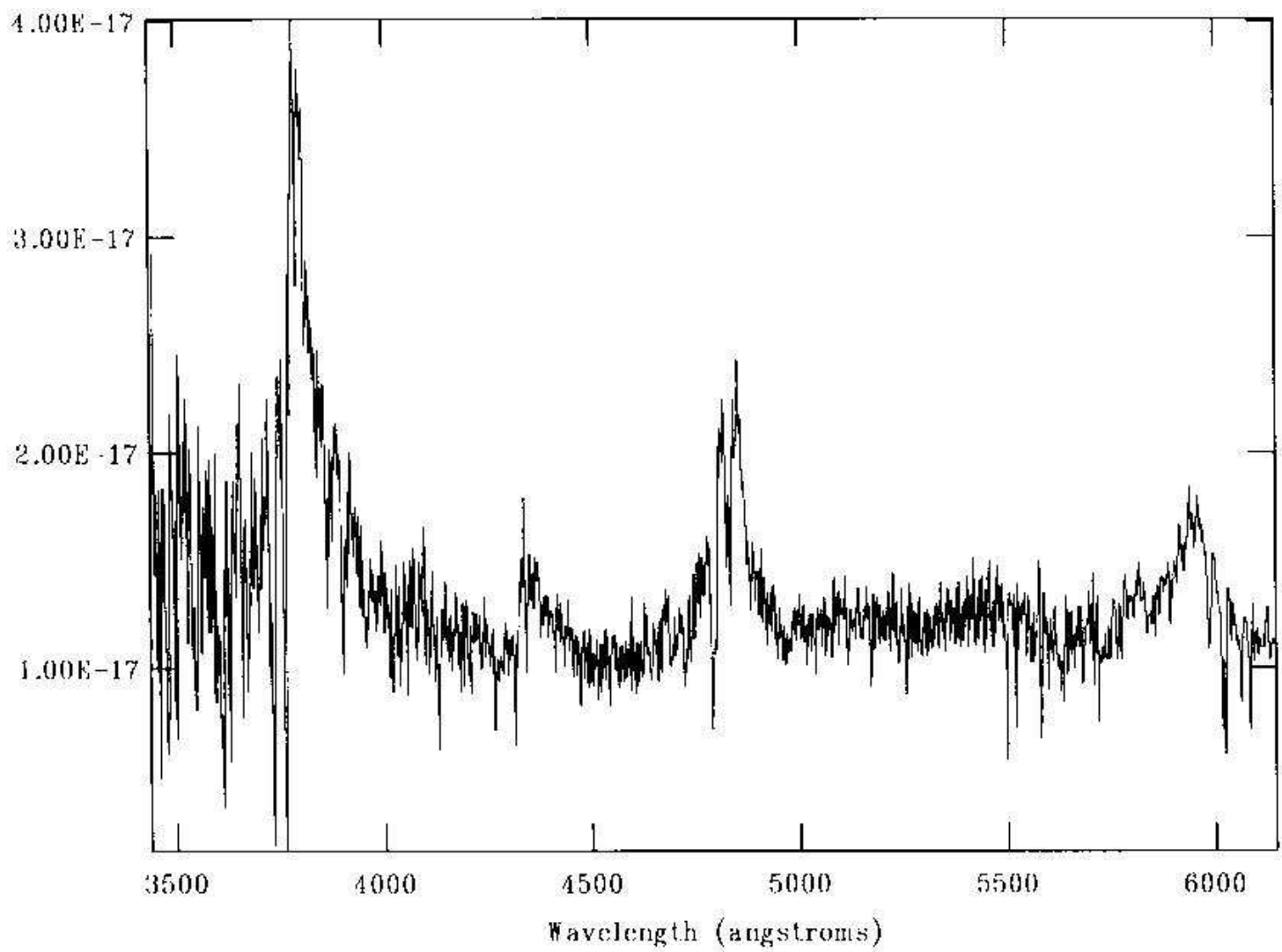
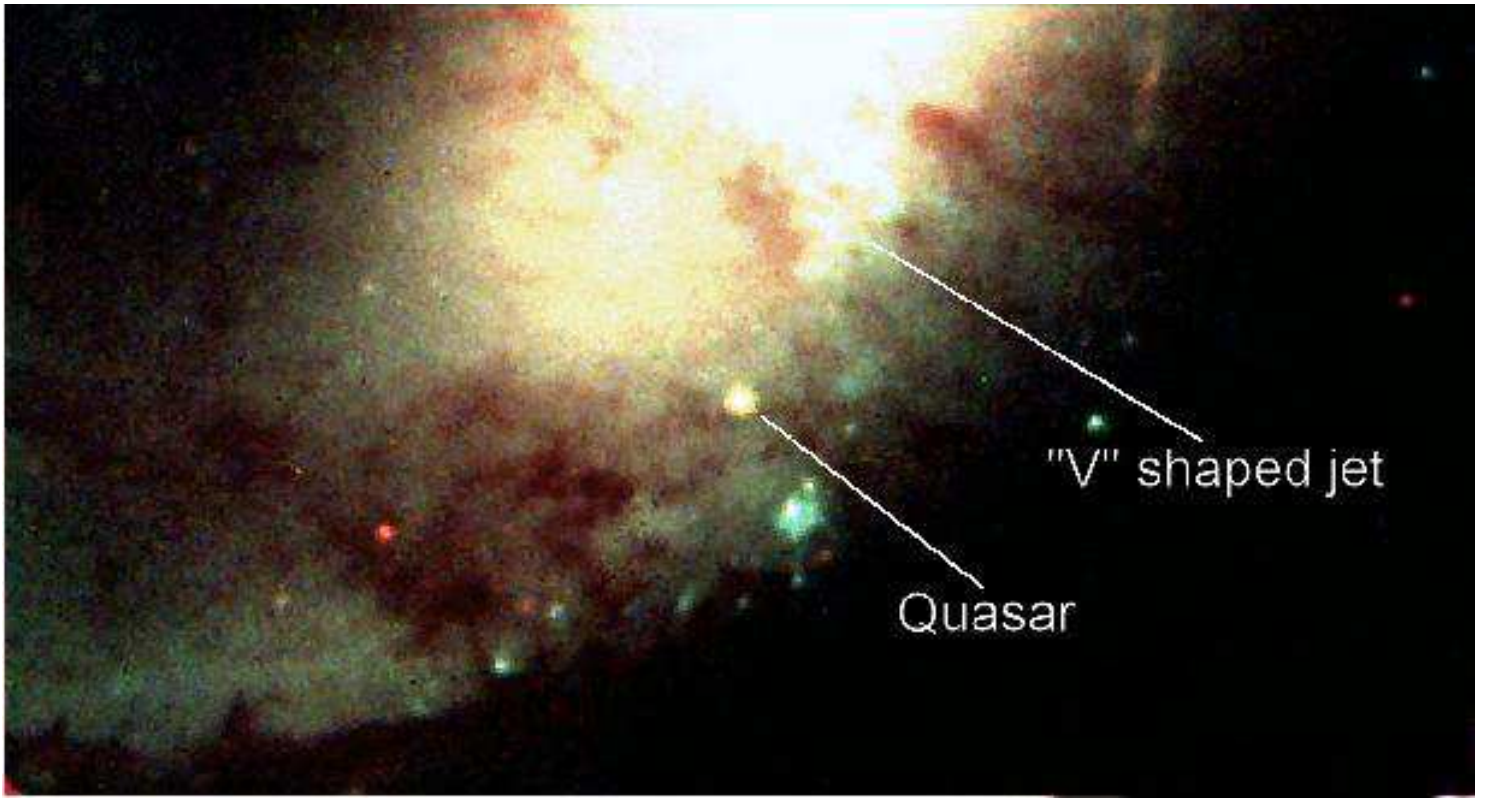
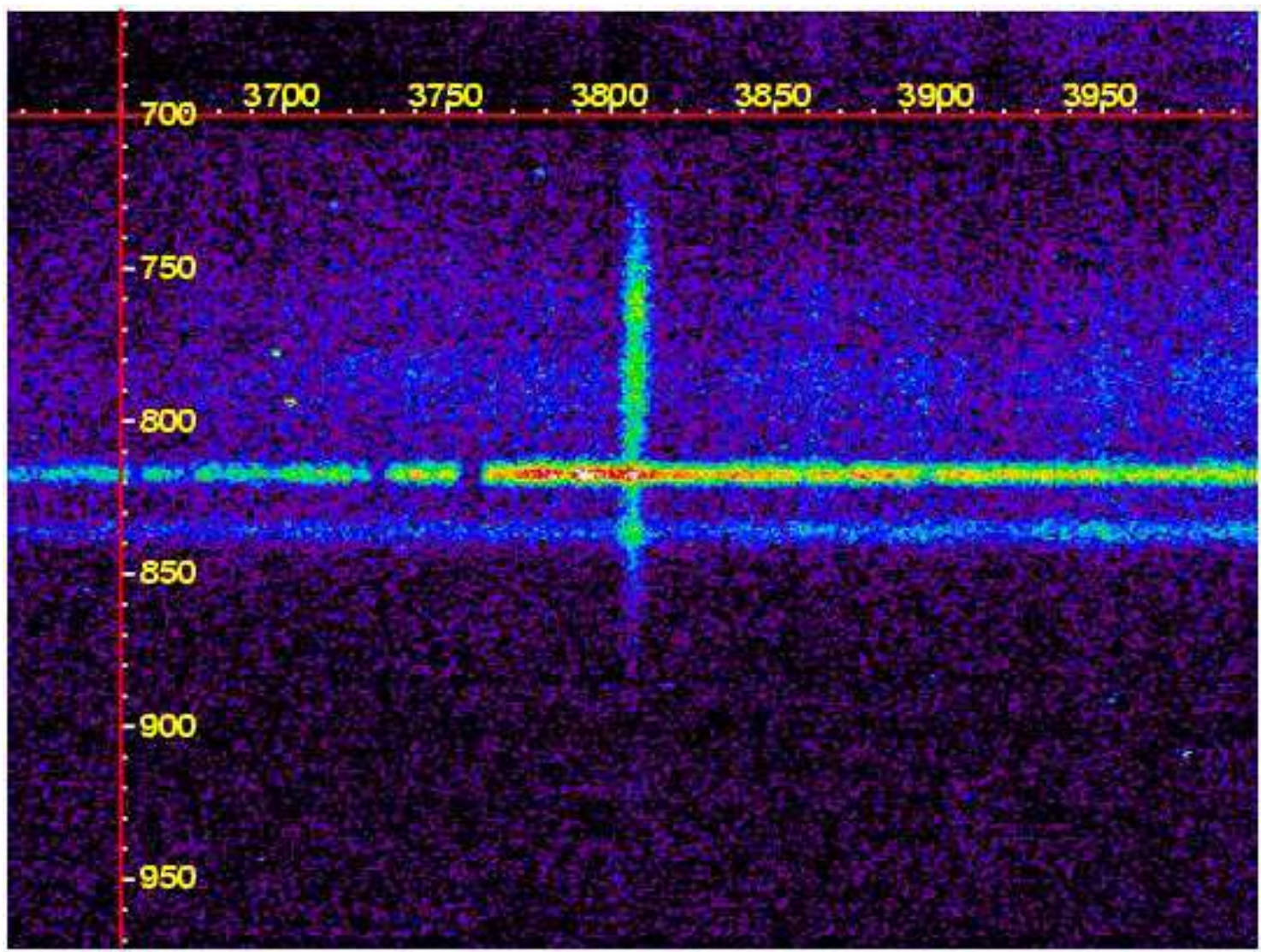
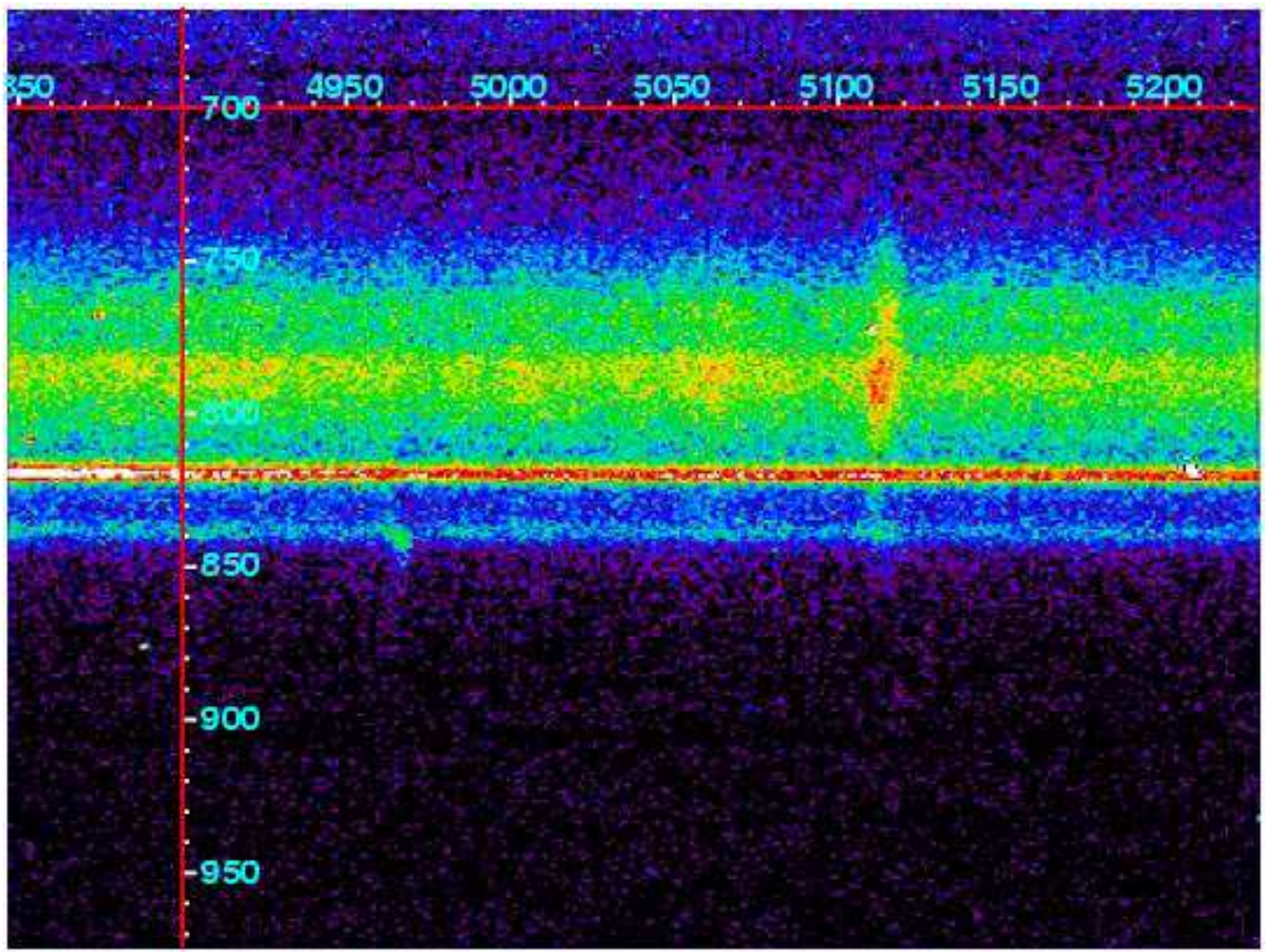
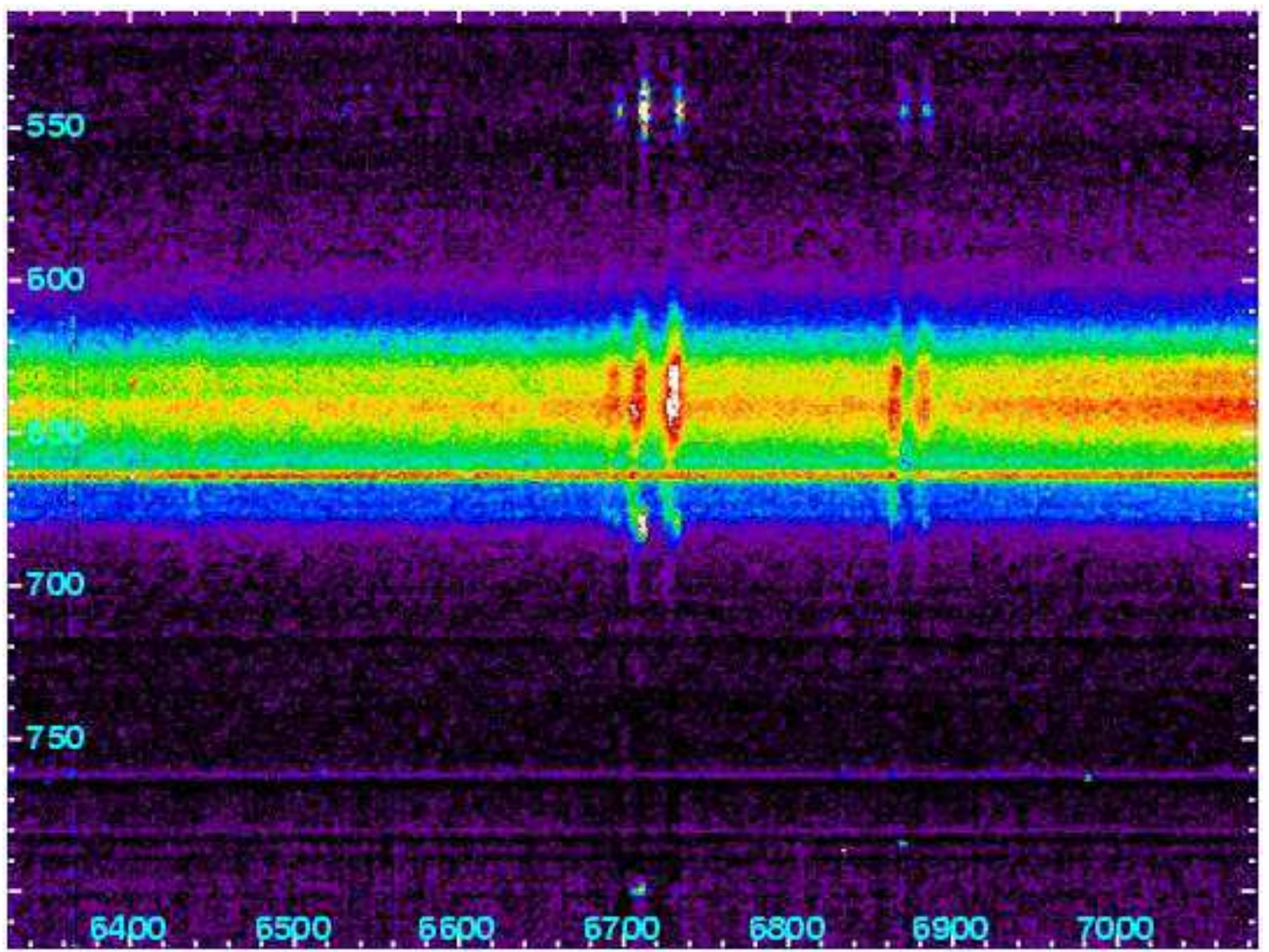
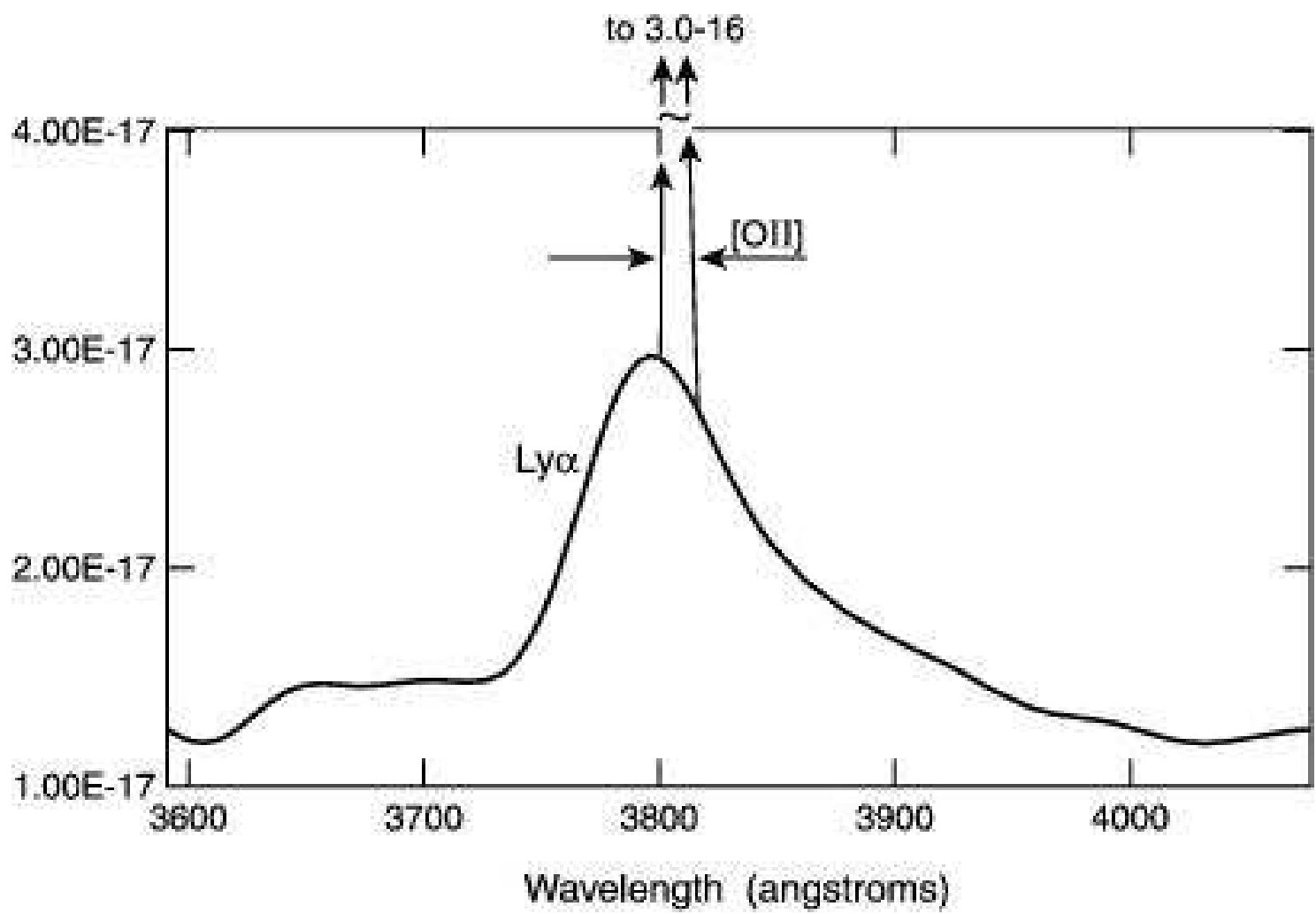
\end{document}